\documentstyle[12pt]{article}

\def\ar{\rightarrow}
\def\bib{\bibitem}

\def\intx{\int\! d^{\sl 4}x}
\def\intX{\int\! d^{\sl 4}X\,}

\def\lar{\longrightarrow}

\def\Lp{L_{_P}}

\def\pa{\partial}

\def\al{\alpha}
\def\be{\beta}
\def\ga{\gamma}
\def\de{\delta}
\def\ep{\varepsilon}

\def\la{\lambda}

\def\si{\sigma}
\def\om{\omega}

\def\Ga{{\it\Gamma}}
\def\La{{\it\Lambda}}
\def\Om{{\it\Omega}}

\def\PSI{{\it\Psi}}

\def\Th{{\it\Theta}}

\def\beq{\begin{equation}}
\def\eeq{\end{equation}}
\def\bed{\begin{displaymath}}
\def\eed{\end{displaymath}}
\def\beqq{\begin{eqnarray}}
\def\eeqq{\end{eqnarray}}
\def\bedd{\begin{eqnarray*}}
\def\eedd{\end{eqnarray*}}

\begin{document}

\centerline{\normalsize\bf II - CONSERVATION OF GRAVITATIONAL ENERGY MOMENTUM} \centerline{\normalsize\bf AND POINCAR\'E-COVARIANT CLASSICAL THEORY OF GRAVITATION}

\vspace*{0.9cm}
\centerline{\footnotesize C. WIESENDANGER}
\baselineskip=12pt
\centerline{\footnotesize\it Aurorastr. 24, CH-8032 Zurich}
\centerline{\footnotesize E-mail: christian.wiesendanger@zuerimail.com}

\vspace*{0.9cm}
\baselineskip=13pt
\abstract{Viewing gravitational energy-momentum $p_G^\mu$ as equal by observation, but different in essence from inertial energy-momentum $p_I^\mu$ naturally leads to the gauge theory of volume-preserving diffeormorphisms of an inner Minkowski space ${\bf M\/}^{\sl 4}$. To extract its physical content the full gauge group is reduced to its Poincar\'e subgroup. The respective Poincar\'e gauge fields, field strengths and Poincar\'e-covariant field equations are obtained and point-particle source currents are derived. The resulting set of non-linear field equations coupled to point matter is solved in first order resulting in Lienard-Wiechert-like potentials for the Poincar\'e fields. After numerical identification of gravitational and inertial energy-momentum Newton's inverse square law for gravity in the static non-relativistic limit is recovered. The Weak Equivalence Principle in this approximation is proven to be valid and spacetime geometry in the presence of Poincar\'e fields is shown to be curved. Finally, the gravitational radiation of an accelerated point particle is calulated.}

\normalsize\baselineskip=15pt

\section{Introduction}

In \cite{chw1} we have started to explore the consequences of viewing the gravitational energy-momentum $p_G^\mu$ as different by its very nature from the inertial energy-momentum $p_I^\mu$, accepting their observed numerical equality as accidential.

As both are conserved this view has led us to look for two different symmetries which through Noether's theorem generate two different conserved four vectors - one obviously being spacetime translation invariance yielding the conserved inertial energy-momentum $p_I^\mu$ vector. To generate an additional conserved four-vector the field concept has proven to be crucial as only fields can carry the necessary inner degrees of freedom to allow for representations of additional inner symmetry groups - in our case an inner translation group yielding the conserved gravitational energy-momentum $p_G^\mu$ vector.

Gauging this inner translation group has then naturally led to the gauge field theory of volume-preserving diffeomorphisms of ${\bf M\/}^{\sl 4}$, at the classical level, thereby generalizing the Yang-Mills approach for compact Lie groups acting on a finite number of inner field degrees of freedom (also see \cite{chwA,chwB} for the mathematical framework). The resulting theory is a consistent classical gauge theory and its gauge fields can be coupled in a universal way to any other field.

For the theory's interpretation there remains the problem of a point-particle limit which also does arise for matter coupled to Yang-Mills theories with any compact gauge group but $U(1)$ where no more than one inner degree of freedom (which is a number) is needed. In Electrodynamics a point-particle can carry charge on its spacetime trajectory and act as a source for the $U(1)$ gauge field defining a consistent theory of point-particle charges and electromagnetic fields as long as we do not ask about the charge back-interacting with the field generated by itself.

In the case of the gauge theory of volume-preserving diffeomorphisms of ${\bf M\/}^{\sl 4}$ coupled to matter the problem of a point-particle limit is aggravated by the occurence of an inner continous space needed to represent the infinite-dimensional gauge group at hands. All fields - both matter and gauge fields - live on the product ${\bf M}^{\sl 4}\times {\bf M}^{\sl 4}$ of two Minkowski spaces. So how can we ever get a point-particle limit?

The solution is related to noting that the full gauge group ${\overline{DIFF}}\,{\bf M}^{\sl 4}$ of volume-preserving diffeomorphisms of ${\bf M}^{\sl 4}$ has a subgroup - the Poincar\'e group $POIN\,{\bf M}^{\sl 4}$ - whose algebra elements display a rather trivial dependence on the coordinates of the inner Minkowski space, namely through the algebra generators only. This in turn allows for an explicit decomposition of all the elements living in the Poincar\'e algebra - i.e. gauge fields, field strengths, currents - into the algebra generators depending on the coordinates of the inner Minkowski space on the one hand and coefficient fields depending on the spacetime coordinates on the other. As a result we get a fully relativistic reduced theory of gauge fields and classical currents carrying a representation of the local inner Poincar\'e group and equations of motion that are covariant under inner Poincar\'e transformations.

This relativistic theory can be solved to leading order in the coupling constant yielding the analogon to Lienard-Wiechert potentials for the Poincar\'e fields. For a specific numerical value of the coupling constant the resulting point particle equation of motion reduces to Newton's laws of gravitation in the non-relativistic limit.

As required, the Weak Principle of Equivalence in the weak field limit is recovered and rods and clocks are shown to behave as predicted by General Relativity (GR) in that limit establishing the relation of the present theory to the geometric modelling of gravitation of GR in the weak field limit. Finally, the dipole radiation of an accelerated point particle is calculated as an application leading to a different prediction than GR does.

The notations and conventions used follow closely those of Steven Weinberg in his classic account on the quantum theory of fields \cite{stw1,stw2}. They are presented in the Appendix.

\section{Local Poincar\'e Invariance, Poincar\'e Gauge Fields and Field Strengths}
In this section we introduce the local Poincar\'e subgroup $POIN\,{\bf M}^{\sl 4}$ of the volume-preserving diffeomorphism group of ${\bf M}^{\sl 4}$ and the corresponding algebra ${\bf poin}\,{\bf M}^{\sl 4}$. We then restrict all relevant elements of the full ${\overline{DIFF}}\,{\bf M}^{\sl 4}$ gauge theory presented in \cite{chw1} to $POIN\,{\bf M}^{\sl 4}$. This allows us to decompose gauge fields, field strenghts, currents etc. living in the gauge algebra into the algebra generators of ${\bf poin}\,{\bf M}^{\sl 4}$ and into coefficient fields depending on the spacetime coordinates. Finally we display the transformation laws of the various fields under local infinitesimal Poincar\'e transformations.

Let us start with the product ${\bf M}^{\sl 4}\times {\bf M}^{\sl 4}$ of two four-dimensional Minkow- ski spaces - one factor representing spacetime with coordinates $x^\mu$, the other factor representing an inner space with coordinates $X^\al$ (all notations and conventions are identical to the ones in \cite{chw1} and are summarized in the Appendix).

This allows us to write local infinitesimal volume-preserving diffeomorphisms of the inner Minkowski space for any fixed spacetime point $x$ as
\beq \label{1}
X^\al (x) \lar X'^\al (x,X) = X^\al (x) + {\cal E}^\al (x,X), \quad \nabla_\al {\cal E}^\al (x,X) = 0, \nonumber
\eeq
where ${\cal E}^\al$ is an infinitesimal differentiable four vector living in inner Minkow- ski space and $\nabla_\al$ acts on inner coordinates $X^\al$ only. The additional condition $\nabla_\al {\cal E}^\al = 0$ ensures volume preservation.

In \cite{chw1} we have developed the gauge theory of ${\overline{DIFF}}\,{\bf M}^{\sl 4}$. Here we will highlight only those aspects of the general theory needed to develop the gauge theory of the local Poincar\'e group $POIN\,{\bf M}^{\sl 4}$, starting with the relevant group representation of ${\overline{DIFF}}\,{\bf M}^{\sl 4}$ on fields $\psi (x,X)$ which are defined on ${\bf M}^{\sl 4}\times {\bf M}^{\sl 4}$.

In the sequel we take the passive view
\beqq \label{2}
x^\mu &\lar& x'^\mu = x^\mu, \quad
X^\al \lar X'^\al = X^\al, \\
\psi (x,X) &\lar& \psi'(x,X) = \psi(x,X) -\, {\cal E}^\al (x,X)\cdot \nabla_\al\,\psi(x,X) \nonumber
\eeqq
in which only the fields are transformed.

The algebra ${\overline{\bf diff}}\,{\bf M}^{\sl 4}$ of this representation is given by the divergence-free gauge parameters ${\cal E}$
\beq \label{3}
{\cal E}(x,X)= {\cal E}^\al (x,X)\cdot \nabla_\al
\eeq
and closes under commutation.

In analogy to non-Abelian gauge theories the local invariance under infinitesimal volume-preserving diffeomorphisms of ${\bf M}^{\sl 4}$ requires the introduction of a covariant derivative and a gauge field
\beq \label{4}
D_\mu (x,X)\equiv \pa_\mu + A_\mu (x,X),
\quad A_\mu (x,X) = A_\mu\,^\al (x,X)\cdot \nabla_\al
\eeq
with components $A_\mu\,^\al$ transforming inhomogenously under local gauge transformations
\beq \label{5}
\de_{_{\cal E}} A_\mu\,^\al = \pa_\mu {\cal E}^\al + A_\mu\,^\be \cdot
\nabla_\be {\cal E}^\al - {\cal E}^\be \cdot \nabla_\be A_\mu\,^\al 
\eeq
to ensure the covariant transformation of $D_\mu$.

Above we have introduced the variation $\de_{_{\cal E}} ..\equiv ..' - ..$ of an expression under a gauge transformation Eqn.(\ref{2}).

The related field strength operator is defined by
\beq \label{6}
F_{\mu\nu} (x,X) = [D_\mu (x,X), D_\nu (x,X)]
\eeq
and its components w.r.t. $\nabla_\al$
\beq \label{7}
F_{\mu\nu}\,^\al = \pa_\mu A_\nu\,^\al - \pa_\nu A_\mu\,^\al 
+ A_\mu\,^\be \cdot \nabla_\be A_\nu\,^\al 
- A_\nu\,^\be \cdot \nabla_\be A_\mu\,^\al 
\eeq
transform homogenously
\beq \label{8}
\de_{_{\cal E}} F_{\mu\nu}\,^\al = F_{\mu\nu}\,^\be \cdot \nabla_\be {\cal E}^\al - {\cal E}^\be \cdot \nabla_\be F_{\mu\nu}\,^\al.
\eeq

It is crucial that $A_\mu$ and $F_{\mu\nu}$ can be decomposed w.r.t $\nabla_\al$. They are elements of the gauge algebra ${\overline{\bf diff}}\,{\bf M}^{\sl 4}$ and their components $A_\mu\,^\al$ and $F_{\mu\nu}\,^\al$ fulfill
\beq \label{9}
\nabla_\al f^\al = 0
\eeq
valid for all algebra elements $f^\al$ ensuring volume preservation. 

There are only two relevant representations of the gauge algebra. First, we have the scalar representation with covariant derivative
\beq \label{10}
D_\mu = \pa_\mu + A_\mu\,^\al\cdot \nabla_\al
\eeq
in which all fields live with the exception of the gauge fields introduced above. In the sequel we will call them matter fields. Second, there is the vector representation with covariant derivative
\beq \label{11}
{\cal D}_\mu^\al\,_\be = \pa_\mu \, \de^\al\,_\be + A_\mu\,^\ga \cdot 
\nabla_\ga \,\de^\al\,_\be - \nabla_\be A_\mu\,^\al
\eeq
in which the gauge fields and field strength components live.

The flat metric $g_{\al\be}$ of the inner Minkowski space in general coordinates transforms as a contravariant tensor \cite{chw1}
\beq \label{12}
\de_{_{\cal E}} g_{\al\be}= - {\cal E}^\ga \cdot \nabla_\ga g_{\al\be} - g_{\ga\be}\cdot \nabla_\al {\cal E}^\ga - g_{\al\ga}\cdot \nabla_\be {\cal E}^\ga.
\eeq

The local Poincar\'e group is now defined by restricting the infinitesimal gauge parameters ${\cal E}$ to those leaving the Minkowski metric $\eta_{\al\be}$ in Cartesian coordinates invariant
\beq \label{13}
\de_{_{\cal E}} \eta_{\al\be}= - \nabla_\al {\cal E}_\be - \nabla_\be {\cal E}_\al = 0\quad {\mbox{or}}\quad \nabla_\al {\cal E}_\be = - \nabla_\be {\cal E}_\al. 
\eeq
Note that the gauge parameters vary with $x$, i.e. are local in spacetime.

The elements $f^\al$ of the corresponding local Poincar\'e algebra are subject to
\beq \label{14}	
\nabla_\al f_\be = -\nabla_\be f_\al, 
\eeq
leaving the metric invariant, and the algebra closes under commutation. For $\nabla_\al {\cal E}_\be = - \nabla_\be {\cal E}_\al $, $\nabla_\al {\cal F}_\be = - \nabla_\be {\cal F}_\al$  we have
\beq \label{15} 
\left[{\cal E}^\al \cdot \nabla_\al, {\cal F}^\be \cdot \nabla_\be \right] = \left( {\cal E}^\al \cdot \nabla_\al {\cal F}^\be 
- {\cal F}^\al \cdot \nabla_\al {\cal E}^\be \right) \nabla_\be
\eeq
with
\beq \label{16} 
\nabla_\al \left({\cal E}^\ga \cdot \nabla_\ga {\cal F}_\be 
- {\cal F}^\ga \cdot \nabla_\ga {\cal E}_\be \right)
= -\nabla_\be \left({\cal E}^\ga \cdot \nabla_\ga {\cal F}_\al 
- {\cal F}^\ga \cdot \nabla_\ga {\cal E}_\al \right).
\eeq

As a result we can immediately take over all formulae of interest from \cite{chw1} by restricting fields living in the gauge algebra ${\overline{\bf diff}}\,{\bf M}^{\sl 4}$ as given in that paper to ${\bf poin}\,{\bf M}^{\sl 4}$ and obtain a gauge theory of the local Poincar\'e group $POIN\,{\bf M}^{\sl 4}$.

In addition, we can explicitly solve the constraints Eqn.(\ref{13}) by setting
\beq \label{17}
{\cal E}^\al (x,X) = \ep^\al (x) + \om^\al\,_\be (x)\, \La^{-1}\, X^\be
\quad {\mbox{with}}\quad \om_{\al\be} (x) = - \om_{\be\al} (x)
\eeq 
which are the familiar expressions for infinitesimal local Poincar\'e transformations where $\ep^\al (x)$ denotes local translations and $\om^\al\,_\be (x)$ local rotations of inner space. The factor $\La^{-1}$ has been introduced so that $\ep$ and $\om$ keep the same inner dimensions of length. As all the gauge algebra elements in ${\bf poin}\,{\bf M}^{\sl 4}$ fulfil the constraints Eqn.(\ref{14}) they can be decomposed in an analogous way. Note that the decomposition allows to make the $X^\al$-dependence of algebra elements explicit leaving us with spacetime dependent coefficients. 

First, we expand the gauge fields
\beqq \label{18}
A_\mu\,^\al (x,X) &=& a_\mu\,^\al (x) + b_\mu\,^\al\,_\be (x)\, \La^{-1}\, X^\be \nonumber \\
{\mbox{with}}\quad b^\mu\,_{\al\be} (x) &=& - b^\mu\,_{\be\al} (x)
\eeqq
calling the coefficients {\it Poincar\'e fields} in the sequel. Note that these fields {\it live on spacetime only}. From Eqn.(\ref{5}) we next read off the inhomogenous transformation behaviour of the $a_\mu\,^\al$ and $b_\mu\,^\al\,_\be$ under gauge transformations Eqn.(\ref{17})
\beqq \label{19}
\de_{_{\ep, \om}} a_\mu\,^\al &=& \pa_\mu \ep^\al + \La^{-1}\, \om^\al \,_\be \, a_\mu\,^\be - \La^{-1}\, b_\mu\,^\al \,_\be \, \ep^\be \nonumber \\
\de_{_{\ep, \om}} b_\mu\,^\al\,_\be &=& \pa_\mu \om^\al\,_\be + \La^{-1}\, \om^\al\,_\ga \, b_\mu\,^\ga\,_\be - \La^{-1}\, b_\mu\,^\al\,_\ga \, \om^\ga\,_\be.
\eeqq

Second, let us expand the field strength components
\beqq \label{20}
F_{\mu\nu}\,^\al (x,X) &=& f_{\mu\nu}\,^\al (x) 
+ h_{\mu\nu}\,^\al\,_\be (x)\, \La^{-1}\, X^\be
\quad {\mbox{with}}\quad \nonumber \\
h^{\mu\nu}\,_{\al\be} (x) &=& - h^{\mu\nu}\,_{\be\al} (x).
\eeqq
From Eqn.(\ref{7}) we then read off the Poincar\'e field strengths in terms of the Poincar\'e gauge fields 
\beqq \label{21}
f_{\mu\nu}\,^\al &=& \pa_\mu a_\nu\,^\al - \pa_\nu a_\mu\,^\al 
+ \La^{-1}\, b_\nu\,^\al\,_\be \, a_\mu\,^\be - \La^{-1}\, b_\mu\,^\al\,_\be \, a_\nu\,^\be \\
h_{\mu\nu}\,^\al\,_\be &=& \pa_\mu b_\nu\,^\al\,_\be - \pa_\nu b_\mu\,^\al\,_\be
+ \La^{-1}\, b_\nu\,^\al\,_\ga \, b_\mu\,^\ga\,_\be 
- \La^{-1}\, b_\mu\,^\al\,_\ga \, b_\nu\,^\ga\,_\be \nonumber 
\eeqq
and from Eqn.(\ref{8}) their variations under a gauge transformation
\beqq \label{22}
\de_{_{\ep, \om}} f_{\mu\nu}\,^\al &=& \La^{-1}\, \om^\al\,_\be \, f_{\mu\nu}\,^\be - \La^{-1}\, h_{\mu\nu}\,^\al\,_\be \, \ep^\be \nonumber \\
\de_{_{\ep, \om}} h_{\mu\nu}\,^\al\,_\be &=& \La^{-1}\, \om^\al\,_\ga \, h_{\mu\nu}\,^\ga\,_\be - \La^{-1}\, h_{\mu\nu}\,^\al\,_\ga \, \om^\ga\,_\be
\eeqq
expressing the expected homogenous transformation behaviour.

\section{Poincar\'e Field Dynamics and Coupling to Classical Point Particles}
In this section we reduce the field equations of the full ${\overline{DIFF}}\,{\bf M}^{\sl 4}$ gauge theory presented in \cite{chw1} to 
field equations for the Poincar\'e fields $a_\mu\,^\al$ and $b_\mu\,^\al \,_\be$ which are defined on spacetime only. Adding a Dirac spinor to the full gauge theory we then heuristically derive classical point-particle currents, the set of classical field equations for the Poincar\'e fields coupled to point particles and the point-particle Lagrangian in the presence of Poincar\'e fields and the respective particle equation of motion.

We start with the action of the full gauge theory
\beq \label{23}
S = -\frac{1}{4\, g^2\, \Lp^2} \intx \intX \Lp^{-4}\, F_{\mu\nu}\,^\al
\cdot F^{\mu\nu}\,_\al + S_M,
\eeq
where we have used the inner scale invariance of the ${\overline{DIFF}}\,{\bf M}^{\sl 4}$ gauge theory discussed in \cite{chw1} to set the parameter $\La$ carrying dimension of length equal to the Planck length $\La = \Lp$.

In natural units with $c=1$, $\hbar =1$, $\Lp = \sqrt \Ga$ equals the square root of Newton's gravitational constant. Note that any other choice of $\La$ is equivalent up to an inner rescaling. In addition we have introduced a dimensionless coupling $g$ which will allow us to count powers in a perturbative expansion. $g$ will be determined at the classical level by comparison with Newton's law for the motion of a point particle in a gravitational field. Finally we have added a generic matter term $S_M$.

Variation of Eqn.(\ref{23}) w.r.t. $A^\nu\,_\al$ yields the field equations \cite{chw1}
\beq \label{24}
\pa^\mu F_{\mu\nu}\,^\al + g\, A^{\mu\,\be}\cdot \nabla_\be F_{\mu\nu}\,^\al 
- g\, F_{\mu\nu}\,^\be\cdot \nabla_\be A^{\mu\,\al} = -\Lp^2\, J_\nu\,^\al,
\eeq
where
\beqq \label{25}
J_\mu\,^\al (x,X) &=& j_\mu\,^\al (x) + k_\mu\,^\al\,_\be (x)\, \Lp^{-1}\, X^\be
\nonumber \\
{\mbox{with}}\quad k^\mu\,_{\al\be} (x) &=& - k^\mu\,_{\be\al} (x)
\eeqq
denote the matter currents $J_\mu\,^\al = \frac{\de S_M}{\de A^\nu\,_\al}$ which can be decomposed into $j_\mu\,^\al$ and $k_\mu\,^\al\,_\be$. Above we have rescaled the gauge fields $A_\mu\,^\al \rightarrow g\, A_\mu\,^\al$ with $g$ which adds additional factors of $g$ in Eqn.(\ref{24}).

Next we decompose the field equations Eqn.(\ref{24}) into reduced equations for the Poincar\'e fields depending on spacetime coordinates only
\beqq \label{26}
\pa^\mu f_{\mu\nu}\,^\al
+ g\, \Lp^{-1}\, a^\mu\,^\be\, h_{\mu\nu}\,^\al\,_\be 
- g\, \Lp^{-1}\, b^\mu\,^\al\,_\be\, f_{\mu\nu}\,^\be &=& -\Lp^2\, j_\nu\,^\al \\
\pa^\mu h_{\mu\nu}\,^\al\,_\be
+ g\, \Lp^{-1}\, b^\mu\,^\ga\,_\be\, h_{\mu\nu}\,^\al\,_\ga  
- g\, \Lp^{-1}\, b^\mu\,^\al\,_\ga \, h_{\mu\nu}\,^\ga\,_\be &=& -\Lp^2\, k_\nu\,^\al\,_\be. \nonumber 
\eeqq

Note that the reduction above has resulted in a relativistic theory with all: the occurence of {\it one} coupling constant, the specific (self-)coupling of the Poincar\'e fields etc. fully determined by $POIN\,{\bf M}^{\sl 4}$ being a subgroup of ${\overline{DIFF}}\,{\bf M}^{\sl 4}$.

In order to determine the classical currents $j_\nu\,^\al$ and $k_\nu\,^\al\,_\be$ let us specify the matter term in Eqn.(\ref{23}) and add a minimally coupled Dirac spinor with action 
\beq \label{27}
S_M = - \intx\intX \Lp^{-4}\,\, {\overline\PSI}\, \left(\ga^\mu (\pa_\mu + g\, A_\mu\,^M\cdot \nabla_M) + m\right) \PSI
\eeq
to the gauge field action.

Note that {\it if the Dirac field would be constant over inner space it would automatically decouple from the gauge fields because gauge fields couple to field variations over inner space only}. We could then integrate over inner space and would recover in effect the free Dirac action we started with before minimal coupling. The same holds true for any other minimally coupled field such as those occuring in the Standard Model action.

Decomposing the gauge field into Poincar\'e fields and focusing on the interaction term in the action Eqn.(\ref{27}) above we have
\beq \label{28}
S_{INT} = \intx\, \left( a_\mu\,^\al\cdot j^\mu\,_\al
+ b_\mu\,^\al\,_\be\cdot k^\mu\,_\al\,^\be \right),
\eeq
where
\beqq \label{29}
j^\mu\,_\al &=& -g\, \intX \Lp^{-4}\,\, {\overline\PSI}\, \ga^\mu\, \nabla_\al \PSI \nonumber \\
&=& -i\, g\, \int\! d^{\sl 4}P\, \Lp^4\,\, \hat{\overline\PSI}\, \ga^\mu\, P_\al \hat\PSI \\
\label{30}
k^\mu\,_\al\,^\be &=& +\frac{g}{2\,\Lp}\, \intX \Lp^{-4}\,\, {\overline\PSI}\, \ga^\mu\, (X_\al \nabla^\be - X_\be \nabla^\al)\PSI \nonumber \\
&=& +\frac{i\, g}{2\,\Lp}\, \int\! d^{\sl 4}P\, \Lp^4\,\, \hat{\overline\PSI}\, \ga^\mu\, M_\al\,^\be \hat\PSI
\eeqq
are the conserved Noether currents expressed in both the inner $X$- and $P$-space which are related to the inner global translation and Lorentz invariances of the action Eqn.(\ref{27}) with generators $P_\al$ and $M_\al\,^\be$ respectively as derived in detail in \cite{chw1}. The corresponding charges generate the inner Poincar\'e algebra which fixes the normalization of the currents in the decomposition above. Note that $j^\mu\,_\al$ carries the dimension of an energy-momentum density and $k^\mu\,_\al\,^\be$ that of an angular-momentum density.

In Electrodynamics to get the classical description of a point charge corresponding to the quantum description of the particle based on a minimally coupled Dirac spinor the Noether current
\beqq \label{31}
j^\mu &=& -i\, e\, {\overline\psi}\, \ga^\mu \psi
= \Big( \mbox{Noether charge}\,\, e = \int d^3 x\, j^{\sl 0} \Big) \nonumber \\
& & \times\, \Big( \mbox{one-particle probability current density}\, -i\, {\overline\psi}\, \ga^\mu \psi \Big) 
\eeqq
of the quantum theory related to global $U(1)$-invariance is replaced by \beqq \label{32}
j^\mu &\ar& \Big( \mbox{Noether charge}\,\, e \Big) 
\times\, \Big( \mbox{point particle current density} \nonumber \\
& &  \int d\tau\, \dot y^\mu\, \de^4 (x-y(\tau))\Big) 
= e\, \int d\tau\, \dot y^\mu\, \de^4 (x-y(\tau))
\eeqq 
with $y^\mu (\tau)$ denoting the classical trajectory of the point particle and $\tau$ the proper time.

In tentative analogy we replace Eqns.(\ref{29}) and (\ref{30}) by the respective Noether charges $P_\al$ and $M_\al\,^\be$ $\times$ the point particle current density
\beqq \label{33}
j^\mu\,_\al &\ar& +g\, \int d\tau\, P_\al(y)\, \dot y^\mu\, \de^4 (x-y(\tau)) \nonumber \\
k^\mu\,_\al\,^\be &\ar& -\frac{g}{2\,\Lp}\, \int d\tau\, M_\al\,^\be(y)\, \dot y^\mu\, \de^4 (x-y(\tau)),
\eeqq
where $P_\al(y)$ and $M_\al\,^\be(y)$ are the inner energy-momentum and angular-momentum respectively carried by a classical point particle along its trajectory $y^\mu (\tau)$. As in Electrodynamics the sign of $g$ has no physical significance and observable expressions will depend on $g^2$ only. But in stark contrast to Electrodynamics where the electric charge appearing in Eqn.(\ref{32}) is a constant both $P_\al(y)$ and $M_\al\,^\be(y)$ vary along the trajectory leading to a much more convoluted situation.

The approximation above is a good one at least for currents Eqns.(\ref{33}) strongly concentrated in three-dimensional space around $y(\tau)$. However, this approximation destroys the local nature of the current appearing in Eqn.(\ref{33}) in inner space and as a consequence reduces the gauge group to the local Lorentz subgroup of the inner Poincar\'e group.

Inserting Eqns.(\ref{33}) into the interaction term Eqn.(\ref{28}) and adding the action of a free relativistic point particle we obtain the action of a point particle in the presence of the Poincar\'e fields 
\beqq \label{34}
S_M &=& -m \int d\tau \\
&+& g \int d\tau\, \left( a_\mu\,^\al (y^\mu)\cdot P_\al(y)
- \frac{1}{2\, \Lp}\, b_\mu\,^\al\,_\be (y^\mu)\cdot M_\al\,^\be(y) \right)\, \dot y^\mu, \nonumber 
\eeqq
which is invariant under local Lorentz rotations
\beqq \label{35}
\de_{_\om}S_M &=& - \frac{1}{2\, \Lp}\, \int d\tau\,\, \pa_\mu
\om^\al\,_\be \cdot M_\al\,^\be \, \dot y^\mu
\nonumber \\
&=& - \frac{1}{2\, \Lp}\, \int d\tau\,\, \frac{d \om^\al\,_\be}{d \tau} 
\cdot M_\al\,^\be = 0 
\eeqq 
if the gauge parameters $\om^\al\,_\be (y(\tau_1)) = \om^\al\,_\be (y(\tau_2)) = 0$ vanish at the boundaries.

Variation of Eqn.(\ref{34}) w.r.t to the particle trajectory yields the relativistic equation of motion of a classical point particle in the presence of Poincar\'e fields
\beq \label{36}
m\, \ddot y_\mu = g\, \left(
f_{\mu\nu}\,^\al (y^\mu)\cdot P_\al(y)
- \frac{1}{2\, \Lp}\, h_{\mu\nu}\,^\al\,_\be (y^\mu)\cdot M_\al\,^\be(y) \right)\, \dot y^\nu,
\eeq
which together with the relativistic field equations for the Poincar\'e fields coupled to the point particle currents Eqn.(\ref{33})
\beqq \label{37}
\pa^\mu f_{\mu\nu}\,^\al
&+& g\, a_\mu\,^\be\, h_{\mu\nu}\,^\al\,_\be 
- g\, b^\mu\,^\al\,_\be\, f_{\mu\nu}\,^\be \nonumber \\
&=& -g\, \Lp^2\, \int d\tau\, P^\al(y)\, \dot y_\nu\, \de^4 (x-y(\tau)) \\
\label{38}
\pa^\mu h_{\mu\nu}\,^\al\,_\be
&+& g\, b^\mu\,^\ga\,_\be\, h_{\mu\nu}\,^\al\,_\ga 
- g\, b^\mu\,^\al\,_\ga \, h_{\mu\nu}\,^\ga\,_\be \nonumber \\
&=& +\frac{g\,\Lp}{2}\, \int d\tau\, M^\al\,_\be(y)\, \dot y_\nu\, \de^4 (x-y(\tau)) 
\eeqq
is the core result of this section.

\section{Leading Order Solution of the Coupled System and Newton's Inverse Square Law}
In this section we solve the coupled system of a classical point particle interacting with the Poincar\'e gauge fields and obtain potentials analogous to the Lienard-Wiechert potentials in Electrodynamics. Identifying the gravitational energy-momentum $P^\mu$ with the inertial energy-momentum $p^\mu$ we obtain Newton's inverse square law for gravity.

In order to extract the physical meaning of the coupled system of Eqns.(\ref{36}) to (\ref{38}) we expand the Poincar\'e fields in powers of $g$ setting
\beq \label{39}
a_\mu\,^\al \equiv \sum_{n = 0}\, g^{2n + 1}\, a^{(n)}_\mu\,^\al,
\quad
b_\mu\,^{\al\be} \equiv \sum_{n = 0}\, g^{2n + 1}\, b^{(n)}_\mu\,^{\al\be}.
\eeq
Inserting the expansions into Eqns.(\ref{37}) and (\ref{38}) we get to leading order $n=0$ the linearized field equations
\beqq \label{40}
\pa^2 a^{(\sl 0)}_\mu\,^\al
&=& - \Lp^2\, \int d\tau\, P^\al(y)\, \dot y_\mu\, \de^4 (x-y(\tau)) \\
\label{41}
\pa^2 b^{(\sl 0)}_\mu\,^{\al\be}
&=& \frac{\Lp}{2}\, \int d\tau\, M^{\al\be}(y)\, \dot y_\mu\, \de^4 (x-y(\tau)), 
\eeqq 
where - using the residual linearized gauge freedom $a^{(\sl 0)}_\mu\,^\al\ar a^{(\sl 0)}_\mu\,^\al + \pa_\mu \ep^\al$ resp. $b^{(\sl 0)}_\mu\,^{\al\be}\ar b^{(\sl 0)}_\mu\,^{\al\be} + \pa_\mu \om^{\al\be}$ leaving $f^{(\sl 0)}_{\mu\nu}\,^\al$ resp. $h^{(\sl 0)}_{\mu\nu}\,^{\al\be}$ invariant - we have chosen the Lorentz gauges $\pa^\mu a^{(\sl 0)}_\mu\,^\al = \pa^\mu b^{(\sl 0)}_\mu\,^{\al\be} = 0$.
 
Eqns.(\ref{40}) and (\ref{41}) are easily solved using the retarded Green function
\beq \label{42}
G_{ret} (x) = -\frac{1}{2\pi}\, \theta(x^{\sl 0})\, \de(x^2)
\eeq
fulfilling $\pa^2 G_{ret} (x-y) = \de^4(x-y)$.

As a result we obtain the Poincar\'e analogues of the Lienard-Wiechert potentials
\beqq \label{43}
a^{(\sl 0)}_\mu\,^\al (x)
&=& \frac{1}{4\pi}\, \Lp^2\, P^\al(y^+)\,
\frac{\dot y^+_\mu}{\dot y^+ \cdot(x - y^+)} \\
\label{44}
b^{(\sl 0)}_\mu\,^{\al\be} (x)
&=& -\frac{1}{8\pi}\, \Lp\, M^{\al\be}(y^+)\, 
\frac{\dot y^+_\mu}{\dot y^+ \cdot(x - y^+)}, 
\eeqq
where we have introduced $y^+ = y(\tau^+)$ which is the uniquely determined retarded point on the particle trajectory with
\beq \label{45}
(x - y^+)^2 = 0,\quad x^{\sl 0} - y^{+\, \sl 0} > 0.
\eeq
For ${\bf r} = {\bf x} - {\bf y}^+$ and $\hat{\bf v} =
\frac{d{\bf y}(\tau^+)}{dt}$ the potentials read more explicitly 
\beqq \label{46}
a^{(\sl 0)}_{\sl 0}\,^\al (x) = \frac{-1}{4\pi}\, \Lp^2\, P^\al\,
\frac{1}{r - {\bf r}\cdot \hat{\bf v}}, \!\!\!\!&&\!\!\!\!
{\bf a}^{(\sl 0)\, \al} (x) = \frac{1}{4\pi}\, \Lp^2\, P^\al\,
\frac{\hat{\bf v}}{r - {\bf r}\cdot \hat{\bf v}}, \\
\label{47}
\!\!\!\!\!\!\!\!\!\!\!\!b^{(\sl 0)}_{\sl 0}\,^{\al\be} (x) 
= \frac{1}{8\pi}\, \Lp\, M^{\al\be}\, 
\frac{1}{r - {\bf r}\cdot \hat{\bf v}}, \!\!\!\!&&\!\!\!\!
{\bf b}^{(\sl 0)\, \al\be} (x) = \frac{-1}{8\pi}\, \Lp\, M^{\al\be}\,
\frac{\hat{\bf v}}{r - {\bf r}\cdot \hat{\bf v}},
\eeqq
where bold letters indicate three-vectors.

In analogy to Electrodynamics we next introduce the Poincar\'e-electric and -magnetic field three vectors d$^{i\,\al}$ and h$^{i\,\al}$ respectively 
\beq \label{48}
\mbox{d}^{i\,\al} \equiv f^{i{\sl 0}\,\al}, \quad
\mbox{h}^{i\,\al} \equiv -\frac{1}{2}\,\ep^{ijk} f_{jk}\,^\al,
\eeq
where $\ep^{ijk}$ is the totally antisymmetric tensor in three dimensions, with a similar decomposition for $h^{\mu\nu}\,_{\al\be}$.

The Poincar\'e-electric and -magnetic fields to leading order are then easily found to be
\beqq \label{49}
& & {\bf d}^\al = -\,\frac{g}{4\pi}\, \Lp^2\, P^\al\,
\frac{({\bf r} - r \hat{\bf v})(1 - {\hat v}^2) + {\bf r}\wedge \left(({\bf r} - r\hat{\bf v})\wedge \frac{1}{c}\, \frac{d\hat{\bf v}}{dt}\right)}
{( r - {\bf r}\cdot \hat{\bf v})^3} \quad\quad\quad\quad \\
& & \label{50} \quad\quad\quad\quad\quad\quad\quad\quad\quad\quad
{\bf h}^\al = \frac{\bf r}{r} \wedge {\bf d}^\al. 
\eeqq
In the static approximation $\hat{\bf v} = 0$ they reduce to
\beq \label{51}
{\bf d}^\al = -\,\frac{g}{4\pi}\, \Lp^2\, P^\al\,
\frac{{\bf r}}{r^3},\quad {\bf h}^\al = 0
\eeq
with analogous expressions for the rotational fields.

The field components are in complete analogy to Electrodynamics and there is radiation by accelerated point particles in the Poincar\'e field theory - terms $\propto \frac{1}{r^2}$ contribute at short distance only and the long-ranging terms $\propto \frac{1}{r}$ in Eqn.(\ref{49}) are the ones related to radiation. The formal analogy of the leading order fields between Poincar\'e field theory and Electrodynamics can be used to easily transfer powerful techniques such as the multipole expansion to the effective field theory.

To further extract the physical content of the theory let us analyze next the motion of a point particle $B$ with trajectory $z^\mu$ and inner energy-momentum $Q^\ga$ and angular-momentum $N^{\ga\de}$ interacting with the fields Eqns.(\ref{49}) and (\ref{50}) generated by another point particle $A$ with trajectory $y^\mu$ and inner energy-momentum $P^\al$ and angular-momentum $M^{\al\be}$ as determined by the equation of motion Eqn.(\ref{36}).

We are specifically interested in the limiting case where $A$ is static and $B$ moves slowly and get from Eqns.(\ref{36}) and (\ref{51}) with $\dot z^\mu \approx (1, \frac{d {\bf z}}{dt})$ for this case
\beq \label{52}
m_{_B}\, \frac{d^2 {\bf z}}{dt^2} = \frac{g^2}{4\pi}\,  
\left( \Lp^2\, P^\al\cdot Q_\al
+ \frac{1}{4}\, M^{\al\be}\cdot N_{\al\be} \right)\,
\frac{{\bf z}}{\mid\! {\bf z}\!\mid^3}.
\eeq

The final and crucial step in order to interpret our theory is the identification of the ordinary and the inner momentum Minkowski spaces ${\bf M}^{\sl 4}\times {\bf M}^{\sl 4}\ar {\bf M}^{\sl 4}$, mapping the inner and inertial energy-momenta and angular-momenta of a point particle onto themselves
\beq \label{53}
P^\al \sim p^\al,\quad M^{\al\be}\sim m^{\al\be}.
\eeq
The physical justification behind this step is the experimental fact that gravitational and inertial masses in the rest frame of a particle or the gravitational and inertial energy-momenta and angular-momenta in an arbitrary frame are the same by measurement (but not by essence in our approach). It is this fact - expressly so far not built into the theory, keeping two sets of energy-momenta and angular-momenta, one participating in the dynamics, one acting as source or experiencing Poincar\'e fields - and the identification Eqn.(\ref{53}) which ultimately allows for a physical interpretation of the theory.

Note that strictly speaking and to keep the logic of calculations correct we should make this identification only after integration of the equation of motion - again because $P^\al$ and $M^{\al\be}$ are not to be confounded with dynamical variables, but measure the coupling strength which depends on the kinematics of the point particle and is not constant as in the case of the motion of a charged point particle in an electromagnetic field.

On this basis we now have for a point particle $A$ fixed at the origin of its rest frame
\beq \label{54}
P^\al = p^\al = (m_{_A},\underline{0}),\quad
M^{\al\be} = m_{\al\be} = 0,
\eeq
and for a point particle $B$ in the same frame
\beq \label{55}
Q^\al = q^\al = (E_{_B}, \underline{q}_{_B}),\quad
E_{_B} = \sqrt{m_{_A}^2 + \underline{q}_{_B}^2}
\eeq 
with $N^{\al\be} = n^{\al\be}$ not vanishing in general.

The bracket in Eqn.(\ref{52}) measures the strength of the interaction and the identification above shows that it is energy-momentum which generates and experiences the Poincar\'e fields. In the rest frame of $A$ the bracket becomes 
\beq \label{56}
\Lp^2\, P^\al\cdot Q_\al
- \frac{1}{2}\, M^{\al\be}\cdot N_{\al\be}
= -\Ga\, m_{_A}\, E_{_B}
\approx -\Ga\, m_{_A}\, \left( m_{_B} + \frac{\underline{q}_{_B}^2}{2\, m_{_B}}\right).
\eeq
Note the crucial minus sign which makes the Poincar\'e force between two point particles - being of tensorial character - attractive which is in stark contrast to a vector theory which always leads to repulsion of charges with the same sign.

To leading order in the rest energy of point particle $B$ we finally recover Newton's inverse square law for gravity
\beq \label{57}
m_{_B}\, \frac{d^2 {\bf z}}{dt^2} = 
-\Ga\, m_{_A}\, m_{_B}\,
\frac{{\bf z}}{\mid\! {\bf z}\!\mid^3}
\eeq
if we fix the coupling constant as
\beq \label{58}
g^2 = 4\pi.
\eeq
Note that this numerical value is dependent on the unit convention chosen for $c, \hbar, \Ga$.

To next to leading order there is a correction to the coupling strength given by the kinetic energy of particle $B$ which in the case of the earth orbiting around the sun amounts to $5\cdot 10^{-7}\,\%$ of the earth's rest mass energy. In addition the earth's orbit is not exactly circular and the coupling strength should vary between the aphelion and the perihelion by about $3.3\,\%$ of the $5\cdot 10^{-7}\,\%$  - producing a potentially measurable effect.

\section{The Weak Principle of Equivalence and Spacetime Geometry in the Poincar\'e Field Theory of Gravitation}
In this section we recover the Weak Principle of Equivalence (WEP) and discuss the impact of the presence of Poincar\'e fields on spacetime measurements.

Let us start with the WEP. As we have built our theory on explicitly keeping the notions of gravitational and inertial energy-momentum and the mechanisms for their conservations strictly separate - not postulating their identity by nature as a foundational principle - we had to account for their observed numerical equality by hand. The WEP should thus be a consequence of the theory at least in the weak field limit.

This is most easily seen starting with the equation of motion Eqn.(\ref{36}) for particle $B$
\beq \label{59}
\ddot y_\mu = g\, \left(
f_{\mu\nu}\,^\al (y^\mu)\cdot \frac{Q_\al}{m_{_B}}
- \frac{1}{2\, \Lp}\, h_{\mu\nu}\,^{\al\be} (y^\mu)\cdot \frac{N_{\al\be}}{m_{_B}} \right)\, \dot y^\nu.
\eeq
We next write the energy-momentum and angular-momentum of $B$ in terms of its inertial mass $m_{_B}$ and its four-velocity $u^\mu_{_B} = \dot y^\mu$
\beq \label{60}
q^\mu = m_{_B}\, u^\mu_{_B},\quad
n^{\mu\nu} = m_{_B}\, (y^\mu\, u^\nu_{_B} - y^\nu\, u^\mu_{_B}),
\eeq
which shows the $m_{_B}$-independence of the r.h.s. of Eqn.(\ref{59}) after identification of gravitational and inertial energy-momentum and angular-momentum. Hence, in this approximation the motion of a point particle in a gravitational field is independent of its mass and the WEP is valid.

Let us turn to the behaviour of rods and clocks in the presence of Poincar\'e fields and analyze the line element which measures gauge-invariant spacetime distances.

The line element has to be the gauge-invariant generalization of the line element at point ${\bf y}$ in the absence of gauge fields
\beq \label{61}
ds^2 = -\, dy_\mu\, dy^\mu = -\, {\dot y}_\mu\, {\dot y}^\mu\, d\si^2,
\eeq 
where $\si$ is an invariant parameter along the spacetime trajectory to be measured and $\dot {}$ denotes a derivative w.r.t. $\si$.

Noting that it is always possible to gauge away Poincar\'e gauge fields in one point \cite{chw1} the natural gauge-invariant generalization of the line element Eqn.(\ref{61}) in the presence of an effective translation background field is
\beqq \label{62}
ds^2 &=& -\, \dot y_\mu\, \dot y^\mu \, d\si^2 \nonumber \\
&+& 2\, g\,
\left( a_\mu\,^\al (y^\mu)\cdot \frac{Q_\al}{m_{_B}}
- \frac{1}{2\, \Lp}\, b_\mu\,^{\al\be} (y^\mu)\cdot \frac{N_{\al\be}}{m_{_B}} \right)\, \dot y^\mu \, d\si^2.
\eeqq
This expression does not depend on the specific test body following the trajectory $y^\mu (\si)$ and applies in particular to any rod or clock on the trajectory - the reason it is indeed measuring invariant spacetime distances.
 
For the gauge fields Eqns.(\ref{46}) and (\ref{47}) generated by a static point particle $A$ with mass $m_{_A}$ located at the origin of its rest mass frame the invariant distance at point $y^\mu$ becomes
\beqq \label{63}
ds^2 &=& \left(1 + \frac{g^2}{2\pi}\,  
\left( \Lp^2\, P^\al\cdot \frac{Q_\al}{m_{_B}}
+ \frac{1}{4}\, M^{\al\be}\cdot \frac{N_{\al\be}}{m_{_B}} \right)\,
\, \frac{1}{r}
\right) dt^2  -\, d{\bf y}^2 \nonumber \\
&\approx& \left(1 -\, 2\, \frac{\Ga\, m_{_A}}{r} \right) dt^2  -\, d{\bf y}^2,
\eeqq
where we have set $g^2 = 4\pi$ in accordance with the Newtonian limit.

This means that clocks in the presence of a static Newtonian potential tick slower or more generally that the spacetime metric measured in experiments is not flat - reproducing a core result of General Relativity \cite{stw4,ros}. Note that this fact does not destroy the fundamental role of the Minkowski metric played in the very development of the Poincar\'e field theory as a theory formulated on flat spacetime \cite{chw1} - essentially because the fundamental fields are the Poincar\'e gauge fields and the measured spacetime metric is just one of many gauge-invariant observables derived from the fundamental fields.

\section{Point-Particle Radiation in the Poincar\'e Field Theory of Gravitation}
In this section we analyze the gravitational radiation of a point particle as an important application of the formalism.

Let us turn to the gravitational radiation of an accelerated point particle. In their linearized form the field equations Eqns.(\ref{36}) to (\ref{38}) can be derived from the action
\beqq \label{64}
S &=& -\frac{1}{4\, \Lp^2} \intx\, f_{\mu\nu}\,^\al
\cdot f^{\mu\nu}\,_\al 
- \frac{1}{4\, \Lp^2} \intx\, h_{\mu\nu}\,^{\al\be}
\cdot h^{\mu\nu}\,_{\al\be} \\
&-& m \int d\tau + g \int d\tau\, \left(
a_\mu\,^\al (y^\mu)\cdot P_\al
- \frac{1}{2\, \Lp}\, b_\mu\,^{\al\be} (y^\mu)\cdot M_{\al\be} \right)\, \dot y^\mu. \nonumber 
\eeqq
Note that this is {\it not} true for the full Poincar\'e field equations, i.e. inserting the various field decompositions such as Eqn.(\ref{18}) into the full action Eqn.(\ref{23}) to get a reduced action Eqn.(\ref{64}) and then varying is not the same as the correct procedure to derive the ${\overline{DIFF}}\,{\bf M}^{\sl 4}$ field equations from the full action Eqn.(\ref{23}) and then to insert the field decompositions to get Eqns.(\ref{36}) to (\ref{38}).

Eqn.(\ref{64}) allows us to immediately write down the gauge-invariant and symmetric energy-momentum tensor to leading order in $g$
\beqq \label{65}
\Th^\mu\,_\nu &=& -\, \frac{1}{4 \Lp^2}\,
\Bigg\{ \frac{1}{4}\,\eta^\mu\,_\nu\, f_{\rho\si}\,^\al \cdot f^{\rho\si}\,_\al - f^{\mu\rho}\,_\al \cdot f_{\rho\nu}\,^\al
\nonumber \\
& & +\, \frac{1}{4}\,\eta^\mu\,_\nu\, h_{\rho\si}\,^{\al\be} \cdot h^{\rho\si}\,_{\al\be} - h^{\mu\rho}\,_{\al\be} \cdot h_{\rho\nu}\,^{\al\be} \Bigg\}.
\eeqq
This tensor is conserved
\beq \label{66}
\pa_\mu\Th^\mu\,_\nu = 0 
\eeq
and the corresponding time-independent momentum four-vector
$ P_\nu = \int\! d^{\sl 3}x\, \Th^{\sl 0}\,_\nu$
fulfills
\beq \label{67}
-\, \pa_{\sl 0} P_\nu = \int\! d^{\sl 3}x\, \pa_i \Th^i\,_\nu.
\eeq

Inserting the Poincar\'e-electric and -magnetic fields Eqns.(\ref{48})  into Eqn.(\ref{65}) for $\nu = 0$ yields the Poynting vector of the theory
\beq \label{68}
\Th^i\,_{\sl 0} = \frac{1}{4 \Lp^2} 
\left( {\bf d}^\al \wedge {\bf h}_\al +  {\bf d}^{\al\be} \wedge {\bf h}_{\al\be} \right)^i.
\eeq
We then obtain from Eqn.(\ref{68}) with the use of Gauss' theorem
\beq \label{69}
-\frac{d P_{\sl 0}}{dt} 
= \frac{1}{4\, \Lp^2}\, \int\! d {\bf S}\, 
\left({\bf d}^\al \wedge {\bf h}_\al +  {\bf d}^{\al\be} \wedge {\bf h}_{\al\be} \right),
\eeq
where $d {\bf S}\equiv {\bf n}\,\, r^2\, d\Om$ is the directed two-dimensional surface element with ${\bf n} \equiv \frac{\bf r}{r}$.

Taking the explicit form of the Poincar\'e-electric and -magnetic fields Eqns.(\ref{49}) and (\ref{50}) in the wave zone $\propto \frac{1}{r}$
\beq \label{70}
{\bf d}^\al = -\,\frac{g}{4\pi}\, \Lp^2\, P^\al\,
\frac{{\bf n}\wedge
\left(({\bf n} - \hat{\bf v})\wedge \frac{d\hat{\bf v}}{dt}\right)}
{( 1 - {\bf n}\cdot \hat{\bf v})^3}\, \frac{1}{r},\quad
{\bf h}^\al = {\bf n} \wedge {\bf d}^\al 
\eeq
we find
\beq \label{71}
{\bf d}^\al \wedge {\bf h}_\al = \left({\bf d}^\al \cdot {\bf d}_\al \right) {\bf n}
\eeq
with an analogous expression for
${\bf d}^{\al\be} \wedge {\bf h}_{\al\be}$.

Defining next
\beq \label{72}
P_{ret} \equiv \frac{d H}{\,\,\,\,dt_{ret}}
= -\frac{d P_{\sl 0}}{\,\,\,\,dt_{ret}},
\eeq
where $H$ denotes the total energy, the radiated angular power distribution is then easily found to be
\beq \label{73}
\!\!\!\! \frac{d P_{ret}}{d\Om}
= \frac{1}{16 \pi}\, 
\left( \Lp^2\, P^\al\cdot P_\al
+ \frac{1}{4}\, M^{\al\be}\cdot M_{\al\be} \right)
\frac{\mid\! {\bf n}\wedge \left(({\bf n} - \hat{\bf v})\wedge
\frac{d \hat{\bf v}}{\,\,\,\,dt_{ret}}\right)\!\!\mid ^2}{( 1 - {\bf n}\cdot \hat{\bf v})^5},
\eeq
where we have used $g^2 = 4\pi$ and where $P^\al$ and $M^{\al\be}$ denote the gravitational energy-momentum and angular-momentum of the radiating point particle respectively. Note that a factor of $\frac{dt}{\,\,\,\,dt_{ret}}= 1 - {\bf n}\cdot \hat{\bf v}$ has decreased the exponent in the denominator by one.

Integrating out the angles leaves us with the relativistic Larmor formula for the linearized Poincar\'e field theory of gravity 
\beq \label{74}
P_{ret} = \frac{1}{6}\, 
\left( \Lp^2\, P^\al\cdot P_\al
+ \frac{1}{4}\, M^{\al\be}\cdot M_{\al\be} \right)\,
\frac{ \mid\!\! \frac{d \hat{\bf v}}{\,\,\,\,dt_{ret}}\!\!\mid ^2 -
\mid\! \hat{\bf v} \wedge \frac{d \hat{\bf v}}{\,\,\,\,dt_{ret}}
\!\!\mid ^2}{( 1 - \hat{\bf v}^2)^3}.
\eeq
Accelerated point particles come along with dipole radiation as they do in Electrodynamics - a prediction which distinguishes the present theory clearly from GR \cite{cmw}.

As an application let us determine the power radiated by the earth along its accelerated orbit around the sun. Approximating the trajectory by a cercle with radius $\rho$, expanding the bracket above in the rest mass $m_{_E}$ of the earth and retaining the leading order term only the formula Eqn.(\ref{74}) further simplifies to
\beq \label{75}
P_{ret} = -\frac{1}{6}\,\Ga\, m_{_E}^2\, 
\frac{\hat{\bf v}^4}{( 1 - \hat{\bf v}^2)^2}
\, \frac{1}{\rho^2}.
\eeq
Putting numbers into Eqn.(\ref{75}) results in a power of $0.5$ Gigawatts radiated off by the earth - giving us not much of a reason to be worried about the earth spiralling into the sun in the near future.

\section{Conclusions}
In this paper we have continued to explore the consequences of viewing the gravitational energy-momentum $p_G^\mu$ as different by its very nature from the inertial energy-momentum $p_I^\mu$.

First we have reduced the gauge theory of volume-preserving diffeomorphisms of ${\bf M}^{\sl 4}$ to a gauge theory of its Poincar\'e subgroup $POIN\,{\bf M}^{\sl 4}$ by expanding all: the gauge fields, field strengths and the diffeomorphism-invariant field equations to obtain reduced gauge fields, field strengths and Poincar\'e-covariant field equations. In contrast to the original fields and equations of motion which depend on both spacetime and inner Minkowski space variables, the reduced fields and equations of motion depend on spacetime variables only and allow for a point-particle limit.

To extract the physical content of the reduced theory we then have derived the Noether current for a Dirac field related to the gauge group and in analogy to the electromagnetic case deduced the Poincar\'e-covariant currents for a classical point-particle. As a result we have obtained a set of non-linear Poincar\'e-covariant field equations coupled to point-particle currents relativistically describing the gravitational fields interacting with matter and matter moving in gravitational fields.

Expanding this set of equations in the coupling and retaining terms to leading order only we then have solved the linear set of equations resulting in Lienard-Wiechert-like potentials for the Poincar\'e fields. After numerical identification of gravitational and inertial energy-momentum and angular-momentum and fixing the numerical value of the coupling constant we finally have recovered Newton's inverse square law for gravity in the static non-relativistic limit.

We finally have shown that the WEP is valid for the reduced Poincar\'e theory and that spacetime geometry in the presence of Poincar\'e fields is curved. As an application we have calculated the radiation of an accelerated point particle in leading order of a multipole expansion resulting in a relativistic Larmor-like formula for the radiated power.

As a consequence of the reduction procedure we have obtained a relativistic description of gravitational fields interacting with matter and of matter moving in gravitational fields - ensuring that the full gauge theory of volume-preserving diffeomorphisms of ${\bf M}^{\sl 4}$ passes all four test criteria set up by Will for a viable theory of gravity \cite{cmw}

{\it (i) A theory of gravity must be complete, i.e. it must be capable of analysing from "first principles" the outcome of any experiment of interest

(ii) It must be self-consistent, i.e. predictions for the outcome of an experiment must not depend on the approach of calculating them if there are different ways

(iii) It must be relativistic, i.e. in the limit as gravity is "turned off" the nongravitational laws of physics must reduce to the laws of special relativity

(iv) It must have the correct Newtonian limit, i.e. in the limit of weak gravitational fields and slow motions, it must reproduce Newton's law.}

Obviously the predictions of the reduced theory such as for radiation of accelerated point particles should not contradict observations and its predictions to next-to-leading order such as for the perihelion precession have yet to be worked out.

Very encouraging though are the facts that the full gauge theory of volume-preserving diffeomorphisms of ${\bf M}^{\sl 4}$ can be quantized consistently, that it is asymptotically free at one loop and that it does look renormalizable to all orders \cite{chw3}.

In summary the following tentative picture is emerging: what we experience macroscopically as gravity is the translation mode of the full gauge field which couples to matter through the Noether current related to inner translation invariance or - for matter being strongly concentrated in small spacetime volumes - through the corresponding Noether charge which plays the role of gravitational energy-momentum. This reproduces Newton's inverse square law for the gravitational force between two point masses without further asumptions. Macroscopically all other degrees of freedom of the full gauge field are frozen down to reasonably small distances or high momenta of the field quanta. Microscopically all modes of the gauge field start contributing and recombine to the full gauge field obeying a quantum dynamics which looks renormalizable as has been explicitly shown at the one loop level in \cite{chw3}.

\appendix

\section{Notations and Conventions}

Generally, ({\bf M}$^{\sl 4}$,\,$\eta$) denotes the four-dimensional Minkowski space with metric $\eta=\mbox{diag}(-1,1,1,1)$, small letters denote spacetime coordinates and parameters and capital letters denote coordinates and parameters in inner space.

Specifically, $x^\la,y^\mu,z^\nu,\dots\,$ denote Cartesian spacetime coordinates. The small Greek indices $\la,\mu,\nu,\dots$ from the middle of the Greek alphabet run over $\sl{0,1,2,3}$. They are raised and lowered with $\eta$, i.e. $x_\mu=\eta_{\mu\nu}\, x^\nu$ etc. and transform covariantly w.r.t. the Lorentz group $SO(\sl{1,3})$. Partial differentiation w.r.t to $x^\mu$ is denoted by $\pa_\mu \equiv \frac{\pa\,\,\,}{\pa x^\mu}$. Small Latin indices $i,j,k,\dots$ generally run over the three spatial coordinates $\sl{1,2,3}$ \cite{stw1}.

$X^\al, Y^\be, Z^\ga,\dots\,$ denote inner coordinates and $g_{\al\be}$ the flat metric in inner space with signature $-,+,+,+$. The metric transforms as a contravariant tensor of Rank 2 w.r.t. ${\overline{DIFF}}\,{\bf M}^{\sl 4}$. Because Riem$(g) = 0$ we can always globally choose Cartesian coordinates and the Minkowski metric $\eta$ which amounts to a partial gauge fixing to Minkowskian gauges. The small Greek indices $\al,\be,\ga,\dots$ from the beginning of the Greek alphabet run again over $\sl{0,1,2,3}$. They are raised and lowered with $g$, i.e. $x_\al=g_{\al\be}\, x^\be$ etc. and transform as vector indices w.r.t. ${\overline{DIFF}}\,{\bf M}^{\sl 4}$. Partial differentiation w.r.t to $X^\al$ is denoted by $\nabla_\al \equiv \frac{\pa\,\,\,}{\pa X^\al}$. 

The same lower and upper indices are summed unless indicated otherwise.

\end{document}